\documentclass[11pt]{article}
\usepackage{graphicx}
\usepackage{amsmath}
\newcommand{\bda}{\begin{\displaymath}\begin{array}{rl}}
\newcommand{\eda}{\end{array}\end{displaymath}}
\newcommand{\beq}{\begin{equation}}
\newcommand{\eeq}{\end{equation}}
\newcommand{\bdm}{\begin{displaymath}}
\newcommand{\edm}{\end{displaymath}}
\newcommand{\bea}{\begin{eqnarray}}
\newcommand{\eea}{\end{eqnarray}}

\newcommand{\ba}{\begin{align}}
\newcommand{\ea}{\end{align}}
\newcommand{\bfig}{\begin{figure}}
\newcommand{\efig}{\end{figure}}

\newcommand{\D}{\displaystyle}

\newcommand{\tin}{t_{\rm in}}

\newcommand{\tplus}{t_{+}}

\newcommand{\omnes}{{\cal{O}}}

\begin{document}

~\vspace{1cm}
\begin{center} 
QCD CORRELATION FUNCTIONS\\ AND THE SHAPE OF $K_{\ell 3}$ FORM FACTORS 

 \vspace{0.7cm}

 Irinel Caprini and Elena-Mirela Babalic

\vspace{0.2cm}
National Institute of Physics and Nuclear Engineering, \\POB MG 6,
Bucharest, R-077125 Romania 

\end{center}

\begin{abstract} 
Bounds on the expansion coefficients of the strangeness changing $K\pi$ form factors
were derived recently from  analyticity and unitarity, using as input suitable correlation functions calculated by perturbative QCD in the Euclidian region. We investigate two types of invariant amplitudes  and their corresponding dispersion relations, and show that they  lead to similar results for the shape of the vector and scalar $K_{\ell 3}$  form factors.
\end{abstract}
%%%%%%%%%%%%%%%%%%%%%%%%%%%%%%%%%%%%%%%%%%%%%%%%%%%%%%%%%%%%%%%%%%%%%%%%%
{\centering\section{ INTRODUCTION\label{sec:intro}}}
%%%%%%%%%%%%%%%%%%%%%%%%%%%%%%%%%%%%%%%%%%%%%%%%%%%%%%%%%%%%%%%%%%%%%%%%
 The semileptonic decays $K\to\pi l\nu_l$ (known also as $K_{\ell 3}$ decays) represent the gold plated channel to extract the element $|V_{us}|$ of the CKM matrix. The decay rate is parametrized in terms of two hadronic form factors  that can not be calculated directly in perturbative QCD. Various methods, including Chiral Perturbation Theory (ChPT), lattice simulations  and dispersion relations  have been considered recently, providing valuable information on the shape of the  form factors in the physical region and increasing the precision of  $|V_{us}|$ extraction (for recent reviews see \cite{Antonelli,KaNe}).

A fruitful approach to the study of the form factors, proposed some time ago  \cite{Okubo,SiRa}, is based on the remark that an upper bound on a weighted integral of the modulus squared of the form factors
along the unitarity cut is sometimes available from independent sources.  This condition can be exploited through complex analysis, leading  to constraints on the values at interior
points or on the  coefficients of the Taylor expansion at zero momentum transfer. Mathematically, the problem belongs to the standard
analytic interpolation theory for the $H^2$ Hardy class of functions, and is  referred to as the Meiman problem \cite{Meiman}.
For the $K_{\ell 3}$ form factors the method was applied in \cite{BoMaRa}-\cite{Abbas:20102}. Recent studies \cite{Abbas:20100}-\cite{Abbas:20102} exploit a modified formalism, proposed in \cite{Caprini2000}, which allows one to include also the phase and modulus of the form factors along the elastic part of the unitarity cut.

The starting point of the approach is a dispersion relation that relates an invariant amplitude calculated by perturbative QCD in the Euclidian region to the hadronic form factors on the unitarity cut. Since the  dispersion relation  requires subtractions, it is convenient to consider derivatives of the invariant functions parametrizing the correlator of two strangeness-changing weak currents. The choice adopted in \cite{Hill}-\cite{Abbas:20102} is suitable especially for the form factors of heavy mesons \cite{LL, BoCaLe}. Alternative definitions of the invariant functions and the corresponding dispersion relations have been investigated in  \cite{BoMaRa} and a comparison of two definitions in the case of the scalar form factor was performed in \cite{BC}. The purpose of the present paper is to compare the predictions of two choices of correlation functions for both the vector and scalar form factors, using the modified version of the formalism, which uses information on the phase and modulus on the elastic part of the unitarity cut.  Also, we use the more precise input available at present, which includes the calculation of the correlators to four loops in perturbative QCD  \cite{KaSt}-\cite{BCH2008} and the recent precise ChPT and lattice calculations \cite{Antonelli, KaNe}.

In section \ref{sec:kl3} we introduce the $K_{\ell 3}$ form factors and in section \ref{sec:qcd} we define the invariant amplitudes of the QCD correlator of the strangeness changing current and the dispersion relations satisfied by them. 
In section \ref{sec:bounds} we briefly review the method of unitarity bounds and in  section \ref{sec:results} we compare the predictions of the two types of dispersion relations for the shape of the vector and scalar form factors.

\vskip0.5cm
%%%%%%%%%%%%%%%%%%%%%%%%%%%%%%%%%%%%%%%%%%%%%%%%%%%%%%%%%%%%%%%%%%%%
{\centering\section{$K_{\ell 3}$ FORM FACTORS\label{sec:kl3}}}
%%%%%%%%%%%%%%%%%%%%%%%%%%%%%%%%%%%%%%%%%%%%%%%%%%%%%%%%%%%%%%%%
 Form factors are defined as usual by  the matrix element 
\begin{eqnarray}
\langle \pi(p_\pi) |V_\mu | K(p_K)\rangle =
(p_\pi+p_K)_\mu\  f _+ (t) + (p_K-p_\pi)_\mu\  f_-^{K \pi } (t)~,        
\label{eq:hadronic element}
\end{eqnarray}
where $V_\mu =\bar{s}\gamma_{\mu}u$ is the hadronic strangeness-changing weak current and  $t=(p_K-p_\pi)^2$ is the squared momentum transfer. 
The vector form factor $f_+(t)$ represents the $P$-wave projection of the crossed channel 
matrix element $ \langle 0 |V_\mu| K \bar\pi \rangle $, whereas the $S$-wave projection
is described by the scalar form factor
\begin{equation}
f _0(t)= f _+(t) + \frac{t}{M_K^2-M_\pi^2} f _-(t).
\label{eq:f0def}
\end{equation}
Several low-energy theorems based on flavour and chiral symmetry are available for the $K_{\ell 3}$ form factors.  At $t=0$, where by construction $f_0(0)=f_+(0)$, $SU(3)$ symmetry implies $f_+(0)=1$. Deviations from this limit
are expected to be small \cite{AdGa} and have been 
calculated in chiral perturbation theory 
and more recently on the lattice  \cite{Antonelli, KaNe}. 
In the case of the scalar form factor, 
current algebra \cite{CallanTreiman} relates the value 
at the Callan-Treiman point
$\Delta_{K\pi}\equiv M_K^2-M_\pi^2$ 
to the ratio  $F_K/F_\pi$ of the decay constants:
\beq\label{eq:CT1}
f_0(\Delta_{K\pi})=\frac{F_K}{F_\pi}  +\Delta_{CT}.
\eeq
To one-loop in ChPT in the isospin limit  $\Delta_{CT}= -3.1\times 10^{-3}$ \cite{GaLe1985}, and the higher order corrections are expected to be small \cite{KaNe}.

Unitarity and causality imply that the form factors $f _+(t)$ and $f _0(t)$ are real analytic functions in the $t$-plane cut along the real axis from the elastic threshold $t_+=(M_K+M_\pi)^2$ to infinity.  According to Watson's theorem \cite{Watson},  below the  
inelastic threshold $\tin$ the phase of each  form factor  is equal (modulo $\pi$) to the 
phase shift of the corresponding partial wave of $\pi K$ elastic scattering. Thus, we can write
\beq\label{eq:watson}
f_+(t+i\epsilon)= |f_+(t)| e^{i\delta_1^{1/2}(t)}, \quad \quad t_+<t< \tin,
\eeq
where $\delta_1^{1/2}(t)$ is the phase shift of the $P$-wave of $\pi K$ elastic scattering with $I=1/2$.  For the scalar form factor a similar relation involving the phase shift  $\delta_0^{1/2}(t)$   of the $S$-wave  holds. The phases $\delta_{1,0}^{1/2}(t)$ are available from the Roy-Steiner analysis of $K\pi$ scattering \cite{BuDeMo} and the dispersive studies of the form factors \cite{Bachirvector}-\cite{Bernard:2009}. Recently,  information on the modulus of the form factors  on the same part of the unitarity cut was  obtained from  the decays  $\tau\to K\pi\nu_\tau$  measured by Belle collaboration \cite{Belle}. 

We mention finally the Taylor expansion  used in most experimental analyses in the physical region $0<t<(M_K-M_\pi)^2$ of $K\to\pi$ semileptonic decays \cite{Lai:2004kb}-\cite{Abouzaid:2009ry}. These expansions are written usually for the ratios $\hat f_{+,0}(t)\equiv  f_{+,0}(t)/f_{+}(0)$ as
\beq
\hat f_+(t)=1 + \frac{\lambda_+'}{M_{\pi}^2}\,t + \frac{\lambda_+''}{2\, M_{\pi}^4}\,t^2+\ldots, \quad\quad
\hat f_0(t)=1 + \frac{\lambda'_0}{M_{\pi}^2}\,t + \frac{\lambda''_0}{2\, M_{\pi}^4}\,t^2+\ldots
       \label{eq:taylor}
\eeq
Stringent bounds on the dimensionless slopes and curvatures $\lambda'$ and $\lambda''$ have been  obtained recently with the method of unitarity bounds \cite{Abbas:20100}-\cite{Abbas:20102}. In the present paper we shall discuss the dependence of these constraints on the dispersion relations for the correlation functions used in the method. 
\vskip0.5cm
%%%%%%%%%%%%%%%%%%%%%%%%%%%%%%%%%%%%%%%%%%%%%%%%%%%%%%%%%%%%%%%%%%%%%%%%%%%%%%%%%%%%%%%%%%%%%%%%
%%%%%%%%%%%%%%%%%%%%%%%%%%%%%%%%%%%%%%%%%%%%%%%%%%%%%%%%%%%%%%%%%%%%%%%%%%%%%%%%%%%%%%%%%%%%%%%%
{\centering\section{QCD CORRELATION FUNCTIONS \label{sec:qcd}}}

%%%%%%%%%%%%%%%%%%%%%%%%%%%%%%%%%%%%%%%%%%%%%%%%%%%%%%%%%%%%%%%%%%%%%%%%%%%%%%%%%%%%%%%%%%%%%%%%
We consider the correlator of two  weak currents $V_\mu$:
\beq\label{eq:ope}
i\int\! d^4x\, e^{iq\cdot x} \langle 0 | T\left\{ V_\mu(x) V_{\nu}(0)^\dagger \right\} | 0 \rangle 
= ( - g_{\mu\nu}q^2+q_\mu q_\nu) \Pi_1(q^2) + q_\mu q_\nu \Pi_0(q^2).
\eeq
According to perturbative QCD, the invariant amplitudes $\Pi_1(q^2)$ and $\Pi_0(q^2)$ satisfy subtracted dispersion relations. It is convenient to write these relations for the functions  $\chi_1(Q^2)$ and $\chi_0(Q^2)$ defined as:
\bea\label{eq:chi1}
\chi_1(Q^2) &\equiv & \frac{1}{ 2} \frac{\partial^2 }{ \partial (Q^2)^2}\left[ Q^2\Pi_1(-Q^2) \right] 
=\frac {1}{ \pi} \int_0^\infty\! dt\,\frac{ t {\rm Im}\Pi_1(t) }{ (t+Q^2)^3 }\,,\\
\chi_0(Q^2)&\equiv & -\frac{\partial}{ \partial Q^2} \left[ Q^2\Pi_0 (-Q^2)\right] 
= \frac{1}{\pi}\int_0^\infty\!dt\, \frac{t {\rm Im}\Pi_0(t)}{ (t+Q^2)^2}\,. \nonumber
\eea
We used here the variable $Q^2=-q^2$, which is positive on the Euclidian axis. 

Alternatively, one can write down convergent  dispersion relations for the slightly different invariant amplitudes $\tilde\chi_1(Q^2)$ and $\tilde\chi_0(Q^2)$: 
\bea\label{eq:chi1tilde}
\tilde\chi_1(Q^2) &\equiv &- Q^2 \frac{\partial \Pi_1 (-Q^2)}{ \partial Q^2} 
= \frac{Q^2 }{ \pi} \int_0^\infty\! dt\,\frac{ {\rm Im}\Pi_1(t) }{ (t+Q^2)^2}\,, \\
\tilde\chi_0(Q^2) &\equiv &\frac{\partial^2 [Q^4 \Pi_0(-Q^2)] }{(\partial Q^2)^2} 
= \frac{2}{ \pi} \int_0^\infty\! dt\,\frac{ t^2 {\rm Im}\Pi_0(t) }{ (t+Q^2)^3}\, .\nonumber  \eea

The functions $\chi_{1,0}(Q^2)$ and  $\tilde\chi_{1,0}(Q^2)$ can be calculated by Operator Product Expansion (OPE) and perturbative  QCD at large $Q^2>0$. At present, calculations  to order 
$ \alpha_s^4$ are available (see \cite{BCH2003}-\cite{BCH2008} and references therein). The perturbation expansions of  $\chi_1(Q^2)$ and $\chi_0(Q^2)$ are 
\bea\label{eq:chi1QCD}
\chi_1(Q^2) = \frac{1}{ 8\pi^2 Q^2}\big[1+ 0.318 \,\alpha_s  -0.062\, \alpha_s^2 
- 0.162\, \alpha_s^3 + 0.176 \,\alpha_s^4 \,\dots \big],\\
\chi_0(Q^2) = 
\frac{3(m_s^2-m_u^2)}{ 8\pi^2 Q^2} \big[ 1 + 1.80\,\alpha_s 
+ 4.65\,\alpha_s^2 + 15.0\,\alpha_s^3 +  57.4 \,\alpha_s^4 \,\dots \big], \nonumber
\eea
while the expansions  of $\tilde\chi_1(Q^2)$ and $\tilde\chi_0(Q^2)$ read
\bea\label{eq:chi1tildeQCD}
\tilde\chi_1(Q^2)&=& \frac{1}{ 4\pi^2}\big[
1+ 0.318 \,\alpha_s + 0.166\,\alpha_s^2
+ 0.205\,\alpha_s^3  + 0.504\, \alpha_s^4  \, \dots \big], \\
\tilde\chi_0(Q^2)& = & 
\frac{3(m_s^2-m_u^2)}{ 8\pi^2 Q^2} \big[1+1.167 \,\alpha_s +1.437 \,\alpha_s^2+ 2.495 \,\alpha_s^3+ 5.254 \,\alpha_s^4\, \ldots\big].\nonumber
\eea
Here $\alpha_s$ is the running QCD coupling and $m_{u,s}$ are the running masses of the quarks at the scale $Q^2$ in $\overline{MS}$ scheme. The higher mass corrections and the condensate contributions are negligible. 

We note that the vector amplitude is renormalization scale invariant, therefore the perturbative expansion of  $\chi_1(Q^2)$ can be obtained from that of $\tilde\chi_1(Q^2)$ by taking the derivative with respect to $Q^2$ and using the renormalization group equation for the running coupling. On the other hand, in the scalar case the derivatives with respect to $Q^2$ must be calculated from the fixed scale perturbative expansions of $\Pi_0$, setting afterwards the scale $\mu^2=Q^2$, since the scalar amplitude is not  renormalization scale invariant\footnote{We are grateful to  Dr. N. Chetyrkin for useful comments on this subject.}.

The connection with the $K_{\ell 3}$ form factors is provided by unitarity: assuming  isospin symmetry, the contribution of the $K\pi$ states to
the positive spectral functions ${\rm Im}\Pi_1(t)$ and  ${\rm Im}\Pi_0(t)$ leads to the inequalities
\bea\label{eq:Pi1}
 {\rm Im}\Pi_1(t) \ge  \frac{3}{ 2}\frac{1}{ 48\pi}\frac{ [(t-t_+)(t-t_-)]^{3/2}}{ t^3} 
|f_+(t)|^2\theta(t-t_+),\\
 {\rm Im} \Pi_0(t) \ge \frac{3}{ 2} \frac{t_+ t_- }{ 16\pi} 
\frac{[(t-t_+)(t-t_-)]^{1/2}}{ t^3} |f_0(t)|^2 \theta(t-t_+), \nonumber\eea
where $t_\pm=(M_K \pm M_\pi)^2$.

By combining the dispersion relations (\ref{eq:chi1}) and  (\ref{eq:chi1tilde}),  with the unitarity relations (\ref{eq:Pi1}), we obtain for each form factor 
two inequalities of the type
\beq
 \int^{\infty}_{\tplus } dt\ \rho(t) |F(t)|^{2} \leq I.
        \label{eq:I}
\eeq
The function $F(t)$ denotes one of the form factors $f_+(t)$ or $f_0(t)$, $\rho(t)\geq 0$ is a positive semi-definite weight function (depending on the momentum $Q^2$)  and $I$ is one of correlation functions (\ref{eq:chi1})-(\ref{eq:chi1tilde}),  calculated at large $Q^2>0$ by perturbative QCD, according to (\ref{eq:chi1QCD}) and (\ref{eq:chi1tildeQCD}). The generic inequality (\ref{eq:I}) is the starting point for deducing analyticity and unitarity constraints on the form factors.

\vskip0.5cm
%%%%%%%%%%%%%%%%%%%%%%%%%%%%%%%%%%%%%%%%%%%%%%%%%%%%%%%%%%%%%%%%%%%%%%%%%%%%%%%%%%%%%%%%%%%%%%%%
{\centering\section{CONSTRAINTS ON THE $K_{\ell 3}$ FORM FACTORS\label{sec:bounds}}}

An overview of the formalism was presented recently in \cite{Abbas:20101}. Here we shall give a summary of the version proposed in \cite{Caprini2000}, which exploits an an optimal way  the information on the phase and modulus for $t_+<t<\tin$, where $\tin$ is the threshold for inelastic channels \cite{Abbas:20100}-\cite{Abbas:20102}.

 We introduce the conformal transformation
\beq\label{eq:ztin}
\tilde z(t) = \frac{\sqrt{\tin}-\sqrt {\tin -t} } {\sqrt {\tin}+\sqrt {\tin -t}}\,,
\eeq
which maps the complex $t$-plane cut for $t>\tin$ onto the unit disc $|z|<1$  in the $z=\tilde z(t)$ plane, and define the 
function $g(z)$  as
\beq\label{eq:gF}
 g(z) = F(\tilde t(z))\, [O(z)]^{-1}\,\omega(z)\, w(z).
\eeq 
Here  $\tilde t(z)$  is the inverse of the function $z=\tilde z(t)$,  $O(z)$ is an Omn\`es function defined by 
\beq	\label{eq:omnes}
O(z) = \omnes(\tilde t(z)),\quad\quad
\omnes (t) = \exp \left(\D\frac {t} {\pi} \int^{\infty}_{\tplus} dt 
\D\frac{\delta (t^\prime)} {t^\prime (t^\prime - t)}\right),
\eeq
in terms of a phase $\delta(t)$, which  is  known from Watson theorem  for 
$t\le \tin$, and is an arbitrary function, sufficiently  smooth ({\em i.e.}
Lipschitz continuous) for $t>\tin$, and 
\beq\label{eq:omega}
 \omega(z) =  \exp \left(\D\frac {\sqrt {\tin - \tilde t(z)}} {\pi} \int^{\infty}_{\tin} {\rm d}t^\prime \D\frac {\ln |\omnes(t^\prime)|}
 {\sqrt {t^\prime - \tin} (t^\prime -\tilde t(z))} \right).
\eeq 
Finally, 
$w(z)$ appearing in (\ref{eq:gF}) is an outer function, {\it i.e.}  a function analytic and without zeros in
$|z|<1$, whose modulus on the boundary ($z=\exp(i\theta)$) is related to the weight $\rho(t)$ appearing in (\ref{eq:I})
and the Jacobian of the transformation (\ref{eq:ztin}) by
 \beq\label{eq:wrho}
\frac{|w(\exp(i\theta))|^2}{2\pi}=  \rho(\tilde t(\exp(i \theta) ))\, \left|\frac{{\rm d} \tilde t(\exp(i \theta))}{{\rm d}\theta}\right|.
\eeq
In the present case the outer functions can be written in a closed analytic form: if we use  the dispersion relations  (\ref{eq:chi1}), the outer functions $w_+(z)$ and  $w_0(z)$ for the vector and scalar form factors, respectively, have the expressions \cite{BoMaRa, BC, Hill}
\bea\label{eq:wchi}
 &&\hspace{-1.5cm}w_+(z)= \frac{1}{8 \sqrt{2 \pi t_{in}}} 
 \sqrt{1-z^2} \\&& \times \ \frac{(1+\tilde{z}(-Q^2))^3 (1-z\, \tilde{z}(t_+))^{3/2} (1-z\, \tilde{z}(t_-))^{3/2}}{(1-z\, \tilde{z}(-Q^2))^3 (1+\tilde{z}(t_+))^{3/2} 
(1+ \tilde{z}(t_-))^{3/2}}, \nonumber\\
&& \hspace{-1.5cm} w_0(z)= \frac{\sqrt{3}(M_K^2-M_\pi^2)}{16 \sqrt{2\pi \tin}}
 \sqrt{1-z}\,(1+z)^{3/2} \nonumber \\
&& \times\,\frac{(1+ \tilde{z}(-Q^2))^2(1-z\,\tilde{z}(t_+))^{1/2}\,(1-z\, \tilde{z}(t_-))^{1/2}}{(1-z\,\tilde{z}(-Q^2))^2(1+ \tilde{z}(t_+))^{1/2}\,
 (1+ \tilde{z}(t_-))^{1/2}}. \nonumber
\eea
Here $z$ is the current variable and $\tilde{z}(t)$ is the  function defined in (\ref{eq:ztin}). 

 The alternative outer functions  $\tilde w_{+,0}(z)$ corresponding to the dispersion relations (\ref{eq:chi1tilde}) differ from (\ref{eq:wchi}) by  simple factors:
\bea\label{eq:wchitilde}
\tilde w_+(z)&=& \sqrt{Q^2}\, \frac{1+\tilde z(-Q^2)}{1- z\,\tilde z(-Q^2)}\, w_+(z) ,\\
\tilde w_0(z)&=&  \sqrt{2}\, \frac{1- z\,\tilde z(-Q^2)}{1+\tilde z(-Q^2)}\, w_0(z). \nonumber
\eea

As proven in \cite{Caprini2000}, the function $g(z)$ defined in (\ref{eq:gF}) is analytic in the unit disc $|z|<1$ and the inequality (\ref{eq:I}) writes as
\beq\label{eq:gI1}
\frac{1}{2 \pi} \int^{2\pi}_{0} {\rm d} \theta |g(\exp(i \theta))|^2 \leq I^\prime,
\eeq
where  
\beq\label{eq:I1}
I^\prime= I - \int^{\tin}_{\tplus} {\rm d}t \rho(t) |F(t)|^2.
\eeq
According to  the so-called Meiman problem, the relation (\ref{eq:gI1}) leads to constraints on the values of $g$ and its derivatives at various points inside the disc $|z|<1$. In the general case, consider the value of $g(z)$ and its first $K-1$ derivatives  at $z=0$, and the values at $N$ other interior points:
\bea\label{eq:cond}
\left[\D \frac{1}{k!} \D \frac{ d^{k}g(z)}{dz^k}\right]_{z=0}&=& g_k, \quad
0\leq k\leq K-1; \nonumber\\
 g(z_n)&=&\xi_n, \quad z_n\ne 0, \quad  1\leq n \leq N, 
\eea
where $g_k$ and $\xi_n$ are given numbers. 
Then the following inequality holds:
\beq\label{eq:det}
\left|
	\begin{array}{c c c c c c}
	\bar{I} & \bar{\xi}_{1} & \bar{\xi}_{2} & \cdots & \bar{\xi}_{N}\\	
	\bar{\xi}_{1} & \D \frac{z^{2K}_{1}}{1-z^{2}_1} & \D
\frac{(z_1z_2)^K}{1-z_1z_2} & \cdots & \D \frac{(z_1z_N)^K}{1-z_1z_N} \\
	\bar{\xi}_{2} & \D \frac{(z_1 z_2)^{K}}{1-z_1 z_2} & 
\D \frac{(z_2)^{2K}}{1-z_2^2} &  \cdots & \D \frac{(z_2z_N)^K}{1-z_2z_N} \\
	\vdots & \vdots & \vdots & \vdots &  \vdots \\
	\bar{\xi}_N & \D \frac{(z_1 z_N)^K}{1-z_1 z_N} & 
\D \frac{(z_2 z_N)^K}{1-z_2 z_N} & \cdots & \D \frac{z_N^{2K}}{1-z_N^2} \\
	\end{array}\right| \ge 0,
\eeq
where
\beq
\bar{\xi}_n = \xi_n - \sum_{k=0}^{K-1}g_k z_n^k, \quad  \quad \bar{I} = I' - \sum_{k = 0}^{K-1} g_k^2.
\eeq
All the  principal minors of the above matrix should also be nonnegative \cite{SiRa,BoMaRa}. An equivalent formulation of the condition (\ref{eq:det}) was presented in \cite{Abbas:20101}.

 The entries of the determinant (\ref{eq:det}) are related, by (\ref{eq:gF}),
to the derivatives $F^{(j)}(0)$, $j\le K-1$  of $F(t)$ at $t=0$, and the values
$F(t(z_n))$. Thus, we obtain from (\ref{eq:det}) constraints on the values of the form factors and their derivatives at $t=0$.

\vskip0.5cm
%%%%%%%%%%%%%%%%%%%%%%%%%%%%%%%%%%%%%%%%%%%%%%%%%%%%%%%%%%%%%%%%%%%%%%%%%%%%%%%%%%%
{\centering\section{RESULTS\label{sec:results}}}
%%%%%%%%%%%%%%%%%%%%%%%%%%%%%%%%%%%%%%%%%%%%%%%%%%%%%%%%%%%%%%%%%%%%%%%%%%%%%%%%%%%
We compare the two types of correlation functions (\ref{eq:chi1}) and  (\ref{eq:chi1tilde}) by presenting the corresponding constraints on the coefficients of the Taylor expansions (\ref{eq:taylor}), which control the shape of the form factors in the semileptonic region. 
In the numerical calculations we took  $Q^2=(2\, \mbox{GeV})^2$ and  $\tin=(1\, \mbox{GeV})^2 $, which is a conservative choice as discussed in Refs. \cite{Abbas:20100}-\cite{Abbas:20102}. The perturbative expansions (\ref{eq:chi1QCD}) and (\ref{eq:chi1tildeQCD}) were evaluated with $\alpha_s(2\, \mbox{GeV})=0.308\pm 0.014$ \cite{Bethke}. The phases $\delta_1^{I=1/2}(t)$ and  $\delta_0^{I=1/2}(t)$ for $t<\tin$ are taken from \cite{BuDeMo}, while above $\tin$ we used smooth phases approaching $\pi$ at infinity.  As proven in \cite{Abbas:20101} and checked numerically with high precision, the results are independent on the parametrization of the phase for $t>\tin$. Finally, we calculated the  low energy integrals in (\ref{eq:I1}) with the parametrizations of $|f_+(t)|$ and $|f_0(t)|$  obtained from the measured  rate of  the $\tau\to K\pi\nu$ decay \cite{Belle}. More details on the input and its uncertainties are given  in \cite{Abbas:20100}-\cite{Abbas:20102}. 

\begin{figure}[htb]
 	\begin{center}\vspace{0.8cm}
 	 \includegraphics[width = 7.cm]{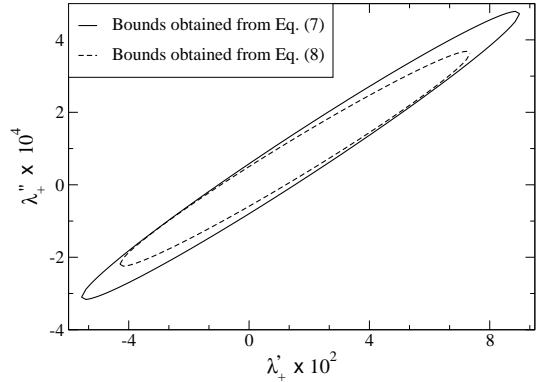}
	\caption{Allowed domains for the slope and curvature of the vector form factor, using as input the correlation function $\chi_1(Q^2)$ from (\ref{eq:chi1}) - solid line,  and the correlation function $\tilde\chi_1(Q^2)$ from (\ref{eq:chi1tilde}) - dashed line. }
	\label{fig:vector}
 	\end{center}\vspace{0.2cm}
\end{figure}

Using the correlation function $\chi_1$ defined in (\ref{eq:chi1}), we obtain from (\ref{eq:det}) the following quadratic inequality for the slope $\lambda_+'$ and the curvature $\lambda_+''$:
\begin{multline}\label{eq:normvect}
f_+^2(0)[(\lambda_+'')^2  - 0.107 \lambda_+' \lambda_+'' + 2.18 \times 10^{-4}\lambda_+''+ 2.98 \times 10^{-3} (\lambda_+')^2\\
- 1.49 \times 10^{-5} \lambda_+'+4.20 \times 10^{-8}] - 4.67 \times 10^{-7}   \leq 0.
\end{multline}
Using alternatively the correlation function $\tilde\chi_1$ defined in (\ref{eq:chi1tilde}), we obtain 
\begin{multline}\label{eq:normvectalt}
f_+^2(0)[(\lambda_+'')^2  - 0.10 \lambda_+' \lambda_+'' + 0.9 \times 10^{-4}\lambda_+''+ 2.59 \times 10^{-3} (\lambda_+')^2\\
- 7.44 \times 10^{-6} \lambda_+'+ 2.64 \times 10^{-8}] - 2.99 \times 10^{-7}   \leq 0.
\end{multline}
These constraints are shown  in Fig. \ref{fig:vector} as the interior of two ellipses in the  slope-curvature plane. For convenience we used the input $f_+(0)=0.962$ quoted in \cite{Antonelli}. One may notice that the domains are very similar, especially in their central regions. 

%%%%%%%%%%%%%%%%%%%%%%%%%%%%%%%%%%%%%%%%%%%%%%%%%%%%%%%%%%%%%%%%%%%%%%%%%%%%%%%%%%%

\begin{figure}[htb]
 	\begin{center}\vspace{0.2cm}
 	 \includegraphics[width = 7.cm]{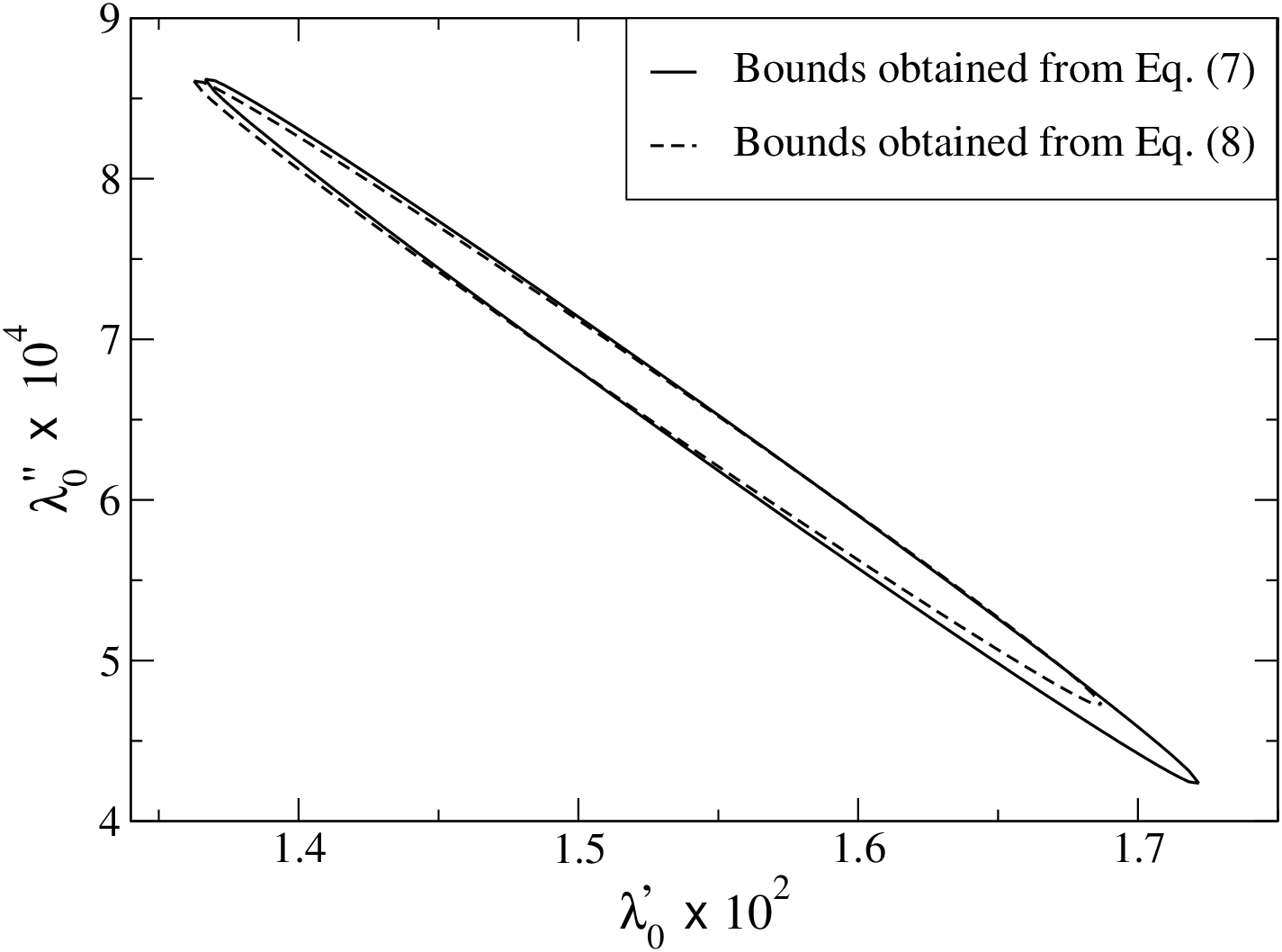}
	\caption{Allowed domains for the slope and curvature of the scalar form factor, using as input the correlation function $\chi_0(Q^2)$ from (\ref{eq:chi1}) - solid line,  and the correlation function $\tilde\chi_0(Q^2)$ from (\ref{eq:chi1tilde}) - dashed line.}
	\label{fig:scalar}
 	\end{center}\vspace{0.2cm}
\end{figure}

For the scalar form factor we shall include an additional input using Callan-Treiman theorem (\ref{eq:CT1}). Using the correlation function $\chi_0$ from (\ref{eq:chi1}), we obtain from (\ref{eq:det}) the condition
\begin{multline}\label{eq:normscCT}
f_+^2(0)\,[(\lambda_0'')^2  + 0.25 \lambda_0' \lambda_0'' + 21.6 \times 10^{-3}\lambda_0'' + 15.3 \times 10^{-3} (\lambda_0')^2 \\
+2.68 \times 10^{-3} \lambda_0'+1.17 \times 10^{-4}] - 10^{-3} f_+(0) f_{CT} 
( 2.67 \lambda_0' \\+ 21.53 \lambda_0''+0.23)
+1.16 \times 10^{-4} f_{CT}^2 
-3.23 \times 10^{-10}   \leq 0.
\end{multline}
where we denoted $f_{CT}=f_0(\Delta_{K\pi})$. The alternative dispersion relation (\ref{eq:chi1tilde}) for the correlation function $\tilde\chi_0$ leads to
\begin{multline}\label{eq:normscCTalt}
f_+^2(0)\,[(\lambda_0'')^2  + 0.24 \lambda_0' \lambda_0'' + 21.1 \times 10^{-3}\lambda_0'' + 14.4 \times 10^{-3} (\lambda_0')^2 \\
+2.52 \times 10^{-3} \lambda_0'+1.12 \times 10^{-4}] - 10^{-3} f_+(0) f_{CT} 
( 2.53 \lambda_0' \\+ 21.02 \lambda_0''+0.22)
+1.11 \times 10^{-4} f_{CT}^2 
-2.64 \times 10^{-10}   \leq 0.
\end{multline}
The numerical coefficients in these constraints are very close, showing that the two types of correlation functions give very similar results. This is illustrated also in Fig. \ref{fig:scalar}, where we represent the domains (\ref{eq:normscCT}) and  (\ref{eq:normscCTalt}) in the  slope-curvature plane for the standard input   $f_+(0)=0.962$ and $f_0(\Delta_{K\pi})= 1.193$.  

%%%%%%%%%%%%%%%%%%%%%%%%%%%%%%%%%%%%%%%%%%%%%%%%%%%%%%%%%%%%%%%%%%%%%%%%%%%%%%%%%%%

\vskip0.5cm
%%%%%%%%%%%%%%%%%%%%%%%%%%%%%%%%%%%%%%%%%%%%%%%%%%%%%%%%%%%%%%%%%%%%%%%%%%%%%%%%%%%
{\centering\section{CONCLUSIONS\label{sec:conclusions}}}
%%%%%%%%%%%%%%%%%%%%%%%%%%%%%%%%%%%%%%%%%%%%%%%%%%%%%%%%%%%%%%%%%%%%%%%%%%%%%%%%%%%
The method of unitarity bounds, especially in the improved version proposed in \cite{Caprini2000}, was shown to be a very useful tool in the phenomenological study of the $K_{\ell 3}$ form factors \cite{Abbas:20100}-\cite{Abbas:20102}.   In the present paper we investigated the dependence of the bounds on the QCD correlation functions and the dispersion relations used as input in the formalism. The results show that two different types of correlation functions lead to similar constraints for the coefficients that control the shape of the vector and scalar form factors at low energies,  providing a nice consistency check of the formalism.

\vskip0.5cm
{\centering\subsubsection*{Acknowledgements}} I.C. thanks B. Ananthanarayan, S. Ramanan, Gauhar Abbas and I. Sentitemsu Imsong for a fruitful  collaboration on the subject of $K_{\ell 3}$ form factors. This work was supported by  CNCSIS in the Program Idei, Contract No. 464/2009,  and by ANCS in the frame of the project PN 09370102.

\vskip0.5cm
%%%%%%%%%%%%%%%%%%%%%%%%%%%%%%%%%%%%%%%%%%%%%%%%%%%%%%%%%%%%%%%%%%%%%%%%%%
\begin{center}

\end{center}
\end{document}